\documentclass{article}

\usepackage{PRIMEarxiv}

\usepackage[utf8]{inputenc} 
\usepackage[T1]{fontenc}    
\usepackage{hyperref}       
\usepackage{url}            
\usepackage{booktabs}       
\usepackage{amsfonts}       
\usepackage{nicefrac}       
\usepackage{microtype}      
\usepackage{lipsum}
\usepackage{fancyhdr}       
\usepackage{graphicx}       
\graphicspath{{media/}}     
\usepackage{adjustbox}
\usepackage{xcolor}
\usepackage{booktabs}
\usepackage{bbm}
\usepackage{comment}
\usepackage{caption}
\usepackage{times}
\usepackage{multirow}
\usepackage{amsmath}
\usepackage{amssymb}
\usepackage{color}

\title{Quantifying the dynamics of peak innovation in scientific careers
}

\author{
  Mingtang Li\\
  Department of Computer Science \\
  University College London \\
  London (UK)\\
  \texttt{mingtang.li.20@ucl.ac.uk} \\
   \And
  Giacomo Livan \\
  Department of Computer Science \\
  University College London \\
  London (UK) \\
  \texttt{g.livan@ucl.ac.uk} \\
   \AND
  Simone Righi \\
  Department of Economics \\
  University Ca' Foscari of Venice \\
  Venezia (Italy) \\
  \texttt{simone.righi@unive.it} \\
}

\begin{document}
\maketitle

\begin{abstract}
We examine the innovation of researchers with long-lived careers in Computer Science and Physics. Despite the epistemological differences between such disciplines, we consistently find that a researcher's most innovative publication occurs earlier than expected if innovation were distributed at random across the sequence of publications in their career, and is accompanied by a peak year in which researchers publish other work which is more innovative than average. Through a series of linear models, we show that the innovation achieved by a researcher during their peak year is higher when it is preceded by a long period of low productivity. These findings are in stark contrast with the dynamics of academic impact, which researchers are incentivised to pursue through high productivity and incremental -- less innovative -- work by the currently prevalent paradigms of scientific evaluation.
\end{abstract}


\section{Introduction}
\label{sec1}
Scientific graduates may choose to pursue an academic career in the hope of innovating their research field. Although a researcher may leave a lasting mark in their field with a single publication, in today's academic environment, building a long-lived career requires a demonstrable ability to produce a steady stream of works, consistently published in reputable peer-reviewed venues. In this respect, a quantitative understanding of scientific innovation ultimately boils down to identifying patterns that may reveal how a researcher's ability to innovate evolves throughout their career, as characterised by the sequence of their publications. Is the ability to innovate roughly constant or does it instead peak at a certain career stage? And, if so, which factors are conducive to periods of high innovativeness in a scientific career?

Answering these questions is especially important in today's highly competitive research ecosystem, characterised by ever-increasing volumes of publications competing for attention~\cite{bornmann2015growth}. In fact, such an environment incentivises researchers to maximise the impact of their scientific work. On the one hand scientific impact is a multifaceted concept, encompassing various dimensions including -- among others -- the plausibility~\cite{Fanelli2009,Fang2012}, originality~\cite{Polanyi2000,Aksnes2019}, scientific value~\cite{Martin1996,Moed2005,Waltman2013}, and societal value~\cite{Lamont2009,Martin2011,Bornmann2012} of scientific publications. On the other hand, however, current academic evaluation practices mostly operationalise scientific impact in terms of bibliometric impact, i.e., the amount of citations that published scientific work receives from other publications~\cite{Amsterdam1995,Radicchi2008,Ellegaard2015}. This, in turn, has led to the proliferation of citation-based bibliometric indicators that seek to quantify different aspects of a paper's or a researcher's ability to attract citations~\cite{Hirsch2005,Radicchi2013,Ioannidis2016}.

There is an abundance of literature showing that equating scientific impact with bibliometric impact shapes the career choices of researchers in at least two ways that are very consequential to scientific innovation. First, it incentivises the pursuit of conservative research in order to publish `safer' contributions~\cite{Fortunato2018,Livan2019}. For instance, it is well known that cross-disciplinary research comparatively attracts fewer citations than incremental research that seeks to develop an already well established field~\cite{levitt2008multidisciplinary,sun2021interdisciplinary}. As much as such an approach to research may be beneficial to the career progression of individual researchers, on a collective level it dramatically stifles risk-taking~\cite{Rzhetsky2015}, which is crucial for scientific progress. Second, it incentivises excessively high productivity (i.e., the so called `publish or perish' culture~\cite{plume2014publish}), resulting in a loss of sustainability in scientific work~\cite{reisz2022loss}. 

One key reason bibliometric indicators of impact are such widely employed is that they easy to use and calculate. The same cannot be said for metrics that quantify innovation. While the study of innovation has a long history in, e.g., Economics~\cite{schumpeter1934theory,schumpeter1939business},  Management~\cite{Dodgson2014}, and Anthropology~\cite{barnett1953innovation}, the development of indicators of innovation is a relatively recent endeavour. Early attempts at quantifying the innovation of scientific publications sought to determine how typical/untypical is the list of references in a paper's bibliography~\cite{Uzzi2013}. However, such measures are highly correlated with interdisciplinarity, and tend to underestimate the innovation of research in well defined fields~\cite{Fontana2020}. An alternative measure -- known as the disruption score~\cite{Wu2019} -- overcomes this issue, and has been shown to be able to reliably discern between innovative and developmental contributions. Its robustness has been validated against data from scientific publications, patents, and software products~\cite{Wu2019,Funk2017}.

Leveraging on this metric, in this paper we seek to quantify the tension between scientific innovation and impact. Specifically, we hypothesise that scientific innovation requires to devote effort to specific projects or research questions over extended periods of time, i.e., ultimately, low productivity. We will seek to validate such hypothesis by tracking the sequence of publications of a large pool of researchers in Computer Science and Physics with long-lived careers. First, we will demonstrate the existence of specific career stretches in which researchers are consistently more innovative than in the rest of their career. Second, we will relate the productivity of a researcher in the run-up to such periods with the level of innovation achieved during them. Our results will show that researchers experience a `magical year' characterised by innovative publications, and that such publications are more innovative when published after a period of low productivity.

\section{Results}
\label{sec2}
We consider researchers in Computer Science and Physics whose career started between 1980 and 2000 and lasted at least $20$ years. Among those, we only retain researchers with at least $10$ published papers and at least one publication every $5$ years (in line with \cite{Li2019}). Overall, this results in a pool of $27,641$ and $34,527$ researchers in Computer Science and Physics, respectively (see Methods). 

We quantify the innovation of researchers in our pool with the disruption score~\cite{Funk2017,Wu2019}, which characterises a paper as more innovative when ensuing publications in the same field cite such a paper more than the publications in its bibliography, i.e., when it eclipses attention to previous work (see Methods). We calculate the disruption score of each paper published by researchers in our pool, and describe the innovation achieved by a researcher throughout their career as the sequence of disruption scores associated to their publications.

\subsection{The non-randomness of scientific innovation}\label{sec21}

We begin our study by measuring how innovation -- as quantified by the disruption score -- evolves throughout a researcher's career. We first identify the year in which a researcher publishes their most innovative paper, i.e., their publication with the highest disruption score. In the following, we shall refer to such a year as the `peak year'. Following the analysis by Sinatra~\emph{et al.} on the randomness of scientific impact during careers~\cite{Sinatra2016}, we then partially randomise a researcher's innovation trajectory across their publication history by keeping the publication dates of their papers intact while randomly reassigning the disruption scores associated to such publications. 

\begin{figure}[h!]
\centering
\includegraphics[width=\textwidth]{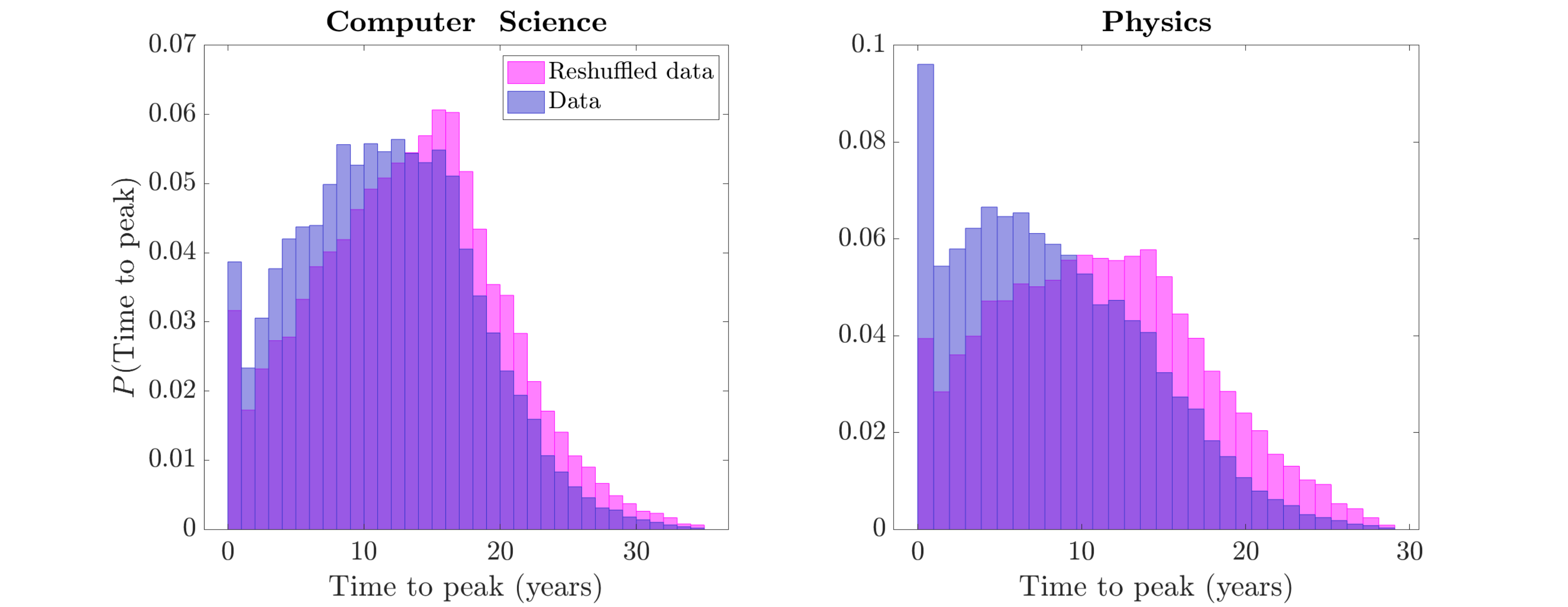}
\caption{Normalised histograms of the time (measured in years) to reach their peak in innovation for researchers in Computer Science (left) and Physics (right) obtained from the original (purple) and randomized (pink) data.}
\label{fig:time_to_peak}
\end{figure}

We calculate the number of years between a researcher's first published paper and their peak year both in the original and in the randomised data. The results are presented as histograms in Fig.~\ref{fig:time_to_peak}. In both disciplines we consider, the distributions obtained from the original and randomised data are significantly different ($p < 0.01$ for both Computer Science and Physics, two-sided Kolmogorov-Smirnov -- KS -- test). We further corroborate this result by comparing the distributions of the number of papers published by researchers before their peak year, which are also significantly different (in both cases $p < 0.01$, two-sided KS test), see Fig.~\ref{fig:papers2peak} in Appendix A.

Based on the above results, we can conclude that, unlike peak impact~\cite{Sinatra2016}, peak innovation does not happen entirely at random in a researcher's career. It is also noticeable that -- on average -- peak innovation happens earlier than one would expect based on our randomised benchmark  ($p < 0.01$ for both Computer Science and Physics, Mann-Whitney U -- MWU -- test).

\subsection{Characterising peak year innovation}

We now proceed to determine whether a researcher's peak year presents peculiar statistical regularities. Specifically, we consider researchers with more than one publication during their peak year (i.e., with at least one more paper published during the peak year in addition to the one with the highest disruption score of their career), and examine whether such papers are also characterised by higher levels of innovation. We hypothesise that the peak year is generally characterised by highly innovative work rather than by just a single innovative paper published among less innovative ones.

To test such conjecture, we split the careers of researchers into a `before peak year' (BPY) and an `after peak year' (APY) phase (we exclude from this analysis researchers whose peak year either occurs during the first or the final year of their career). We then compute the average level of innovation achieved by the researchers in our pool in both phases (for BPY, $N_\mathrm{CS} = 24,209$ in Computer Science and $N_\mathrm{PHY} = 31,139$ in Physics; for APY, $N_\mathrm{CS} = 25,781$ and $N_\mathrm{PHY} = 33,465$). We then compute an average `peak year' (PY) innovation level from the papers published during such year, excluding the paper responsible for the peak itself (for PY, $N_\mathrm{CS} = 18,640$ and $N_\mathrm{PHY} = 26,543$).

\begin{figure}[h!]
\centering
\includegraphics[width=\textwidth]{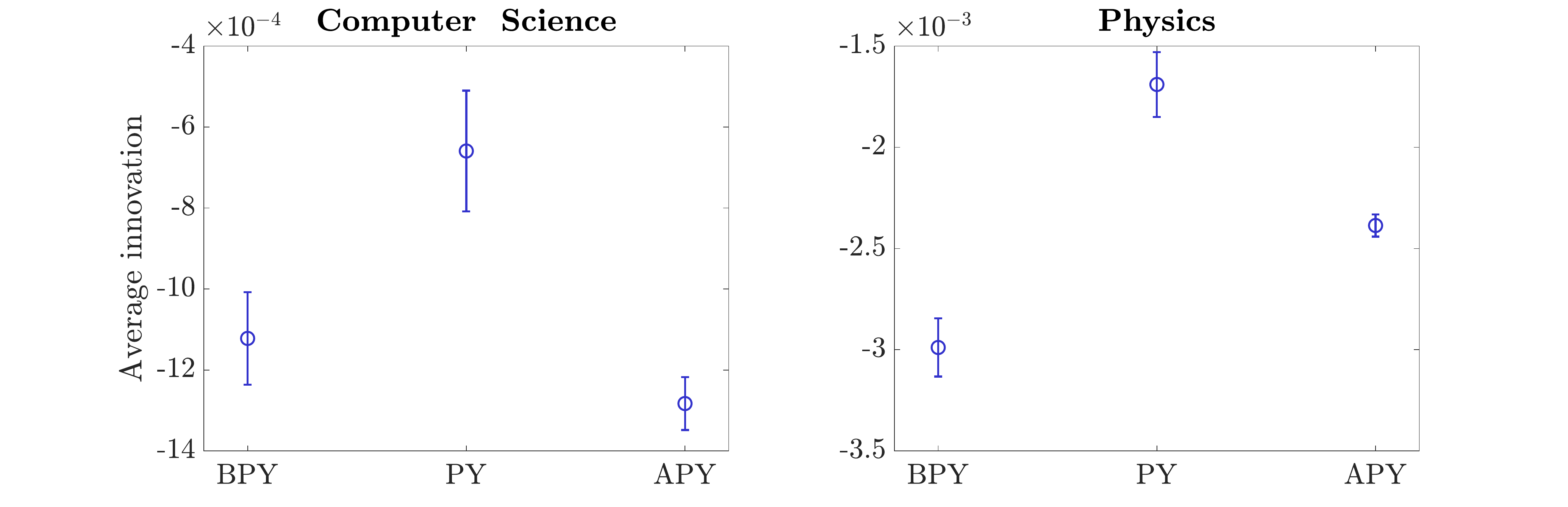}
\caption{Average innovation levels (disruption scores) achieved by researchers with papers published before their peak year (BPY), during their peak year (excluding the paper responsible for the peak itself, PY) and after their peak year (APY). Error bars represent standard errors.}
\label{fig:before_after_peak}
\end{figure}

By comparing innovation levels in the three groups, we find that the distributions of disruption scores in the BPY and APY phases are significantly different from those in the peak year ($p < 0.01$ in both cases, two-sided KS test). More importantly, we find that the average disruption score is higher in the latter year than in the other two phases ($p < 0.01$ in both cases, MWU test), see Fig.~\ref{fig:before_after_peak}.

In order to mitigate potential biases due to differences in length of the before/after peak phases, we repeat the above tests with before/after peak year phases of 2 years (for BPY, $N_\mathrm{CS} = 20,980$ and $N_\mathrm{PHY} = 28,867$; for APY, $N_\mathrm{CS} = 21,987$ and $N_\mathrm{PHY} = 30,792$), finding equivalent results (except for $p = 0.021$ in the MWU test between PY and BPY in Computer Science), see Fig.~\ref{fig:innovation_levels_2yrs} in Appendix B.

\subsection{Determinants of peak-year innovation}

Having established the existence of a `magical year' characterised by generally higher levels of innovation, and having established that it does not occur at random in a researcher's career, we now seek to establish its relationship with the effort devoted to achieve that peak. In order to do so, we quantify the overall innovation of a researcher at their peak year as the average disruption score of their papers published during such year, including the paper responsible for the peak in innovation.

We calibrate a series of linear models to investigate the relationship between the level of innovation achieved by a researcher in their peak year and a set of variables aimed at quantifying the effort put into the scientific work published during that year. We aim to capture the tradeoff between the incentive to publish frequently -- i.e., productivity -- and the ability to produce innovative contributions. We expect that the more time is spent working on a paper, the more likely it is to be an innovative one. We quantify effort as the average time (measured in years) spent to work on a paper published during a period of interest. More specifically, we consider the ratio between a variable we shall refer to as `time devoted' and the number of papers published by a researcher (i.e., their productivity) during the period of interest. The former is defined as the number of years between a researcher's last publication before the period of interest and the final year of that period. For instance, the effort in 2007-2008 of a researcher who published 1 paper in 2005, 3 in 2007, and 5 in 2008 would be 3/8.

We also define corresponding relative quantities, i.e., measures of effort, time devoted, and productivity computed as the ratio between such quantities during a period of interest and over the entire career of a researcher. In Appendix C Fig.~\ref{fig:distributions}, we plot the distributions of such quantities when using the peak year as the period of interest. In the following, we only present results obtained based on such relative quantities, given that the results obtained from the corresponding absolute quantities are qualitatively equivalent (see Appendix D Tables 1-4).

In three distinct models, we measure the linear relationship between a researcher's innovation during their peak year and the relative effort, relative productivity, and relative time devoted associated with that year. In each model we control for the same additional factors. Indeed, we include the time to reach the peak year (measured in years) and the calendar year corresponding to the peak year in order to control for possible temporal effects. We also include the researcher's number of coauthors in their papers published during the peak year to control for network effects. Lastly, we include the average disruption score of papers published in the two years before the peak year to control for idiosyncrasies in the ability to innovate. As a robustness check, we also calibrate models with the above variables computed in the two years before the peak as additional controls. The results we obtain in these models are qualitatively equivalent to the main results presented here (see Appendix D Tables 5-8).

\begin{figure}[h!]
\centering
\includegraphics[width=\textwidth]{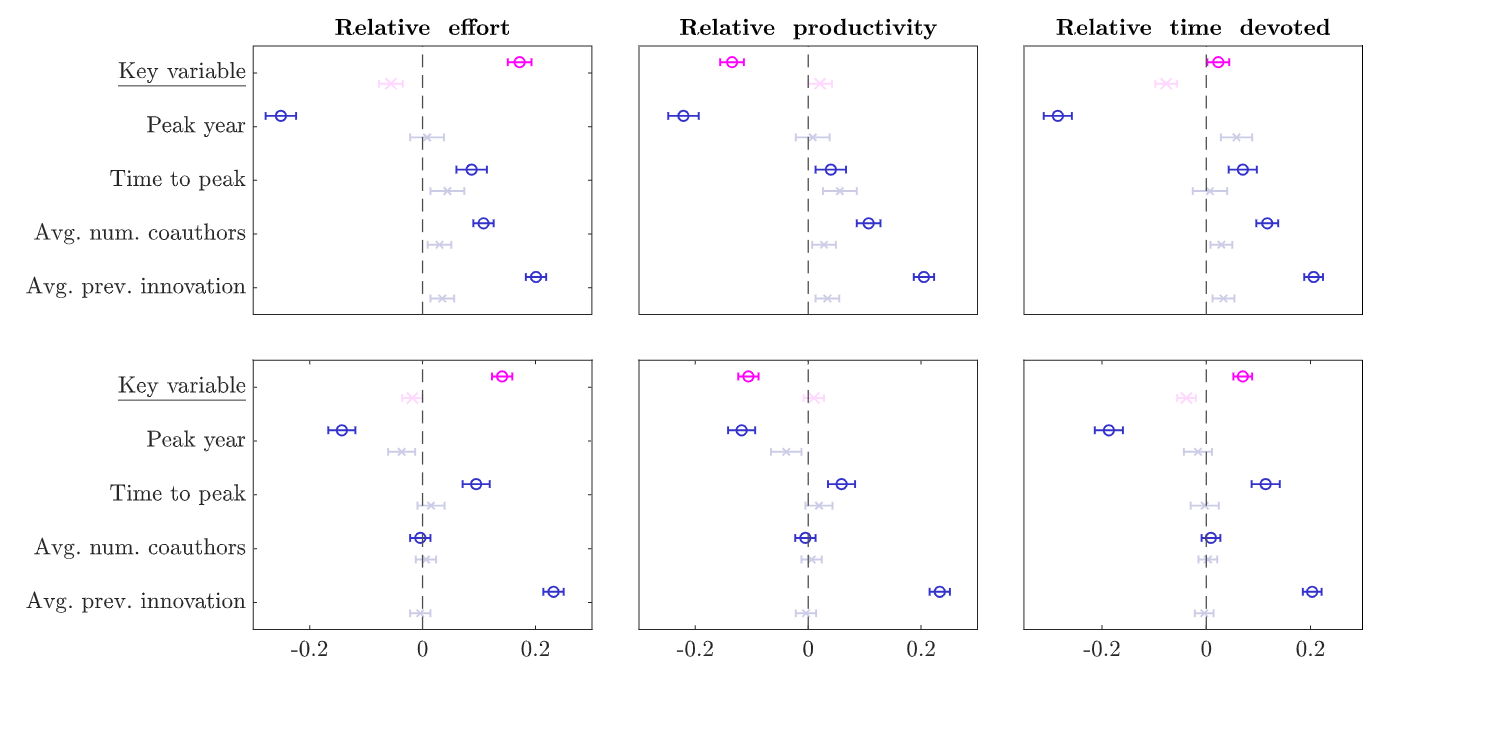}
\caption{Coefficients of linear regression models for peak year innovation (circles, bright colours) and impact (crosses, dim colours). Each column displays results obtained with a different key variable of our analysis, i.e., relative effort (left), relative productivity (center), relative time devoted (right). Error bars represent standard errors multiplied by three. Models in the top (bottom) row refer to Computer Science (Physics). From left to right, in Computer Science models for innovation have an $R^2$ of $0.142$, $0.132$ and $0.116$, while models for impact have an $R^2$ of $0.009$, $0.007$ and $0.011$. In Physics, models for innovation have an $R^2$ of $0.086$, $0.078$ and $0.071$, while models for impact have an $R^2$ of $0.001$, $0.001$ and $0.002$. See Appendix D Tables 1--4 for detailed regression tables.}
\label{fig:regression}
\end{figure}

The results of the aforementioned models are illustrated in Fig.~\ref{fig:regression}. Both in Computer Science and Physics relative effort contributes positively to the innovation level achieved during the peak year, and the same holds for relative time devoted. Contrastingly, relative productivity is found to be detrimental to peak year innovation. Overall, we interpret these results as a clear sign that peak innovation improves when preceded by a period of focus on the work that will be eventually published during the peak year, and working on less papers around the peak years yields higher peak innovativeness. 

A priori, one could not rule out that innovation measured with the disruption score provides similar information to the impact of a paper. In order to test this hypothesis, we calibrate the linear models described above with the same set of independent variables and impact as dependent variable. We measure the latter with the accumulated citations of peak year papers over the first 5 years after their publication. We find the explanatory power of such models to be quite low. Yet, they still provide important insight on the correlations between our effort-related variables and scientific impact. In fact, we find that higher relative productivity contributes to a higher impact, whereas relative effort and relative time devoted have the opposite effect (see Fig.~\ref{fig:regression}). Similar considerations apply when considering the corresponding absolute quantities (See Appendix D Tables 1-8).

The control variables in our model also contribute to explain the determinants of peak scientific innovation. Indeed, we find that researchers who reached their peak in more recent years achieved -- on average -- lower innovation levels, in line with various studies that have evidenced how science has become less innovative over time~\cite{JONES2009,Buchanan2015,Youn2015,Gold2021}. After controlling for such an effect, we observe a positive relationship between peak innovation and the time to reach the peak.

Our results also show that -- in both disciplines -- the more innovative a researcher's previous papers are, the more innovative they will be in the peak year. Finally, we observe that publishing with more co-authors has a positive impact on the peak year innovation level in Computer Science, but not in Physics.

\section{Discussion}

We examined the career dynamics of researchers from the perspective of their innovation, defined in terms of disruption scores~\cite{Wu2019,Funk2017}. We find that peak innovation does not occur entirely at random in a researcher's career, and that the time it takes for most researchers to reach their peaks varies depending on their discipline. This is in contrast with the so called random impact rule~\cite{Sinatra2016,Liu2018}, which states that each paper in a researcher's career is equally likely to become their most impactful one (i.e., their most cited publication). In particular, we show that innovation peaks in a scientific careers happen -- on average -- earlier than one would expect based on a `random innovation rule', i.e., a null model that randomises innovation across the sequence of publications in a researcher's career.

Our findings resonate with observations that younger researchers tend to be more innovative than more senior ones~\cite{packalen2019age,diamond1980age} due, e.g., to more innovative individuals being selected-out of academia \cite{kim2003impact}, weakened incentives to innovate after tenure \cite{diamond1980age}, or cognitive decline \cite{dietrich2007optimal}. At the same time, our regression analysis shows that innovation peaks that happen later in a career tend to be higher. This suggests that the intuition that senior scientists tend to be less innovative needs to be qualified. While it is rarer than expected to observe a researcher achieving their peak later in their career, when this happens it tends to lead to higher innovation (after controlling for the covariates included in our analysis). We speculate that this may be related to the overall decline of innovation in science~\cite{JONES2009,Buchanan2015,Youn2015,Gold2021} -- which we consistently detect in our models -- which may leave more room for experienced researchers to innovate.

Research on the dynamics of careers shows that individuals tend to experience `hot streak' periods in which they enjoy sustained success in their work~\cite{Liu2018,williams2019quantifying,Liu2021}. Our results demonstrate that a similar effect holds for scientific innovation. In fact, we show that the peak year is characterised not just by one very innovative paper, but by a series of papers that are more innovative than average. In other words, our findings show that papers published during the peak year enjoy a synergistic effect, i.e., the paper responsible for the peak does not drain a researcher's resources away from other publications.

Having determined that peak productivity does not happen at random and it is part of a special period of high innovativeness, it is then important to shine light on the determinants of the magnitude of the peak. To this end, we run a series of regression models where the magnitude of the peak in innovation is regressed against various measures that proxy the effort infused by an individual in their peak year papers and various measures of productivity. We find that devoting more time and effort to the research published in the peak year has a positive correlation with the innovation level achieved during it, while measures of productivity have a negative correlation with it. One potential explanation for this result is that working on many papers at the same time may drain researchers of energy and ideas. Notably, splitting findings from the same studies into multiple papers is often a deliberate strategy -- often referred to as `salami slicing'~\cite{jackson2014multiple} -- aimed at maximising impact in terms of expected citation volumes. 

Such a tension between innovation and impact is well captured by our models, which show that those two dimensions have opposite correlations with the aforementioned proxies of effort and productivity. Innovation thrives when more effort is infused into scientific work, i.e., experiencing periods of low productivity. Conversely, low productivity is detrimental to impact. While one should not read too much into the latter result due to the reduced amount of variation explained by our models for impact, these findings strongly suggest that scientific innovation and impact follow very different patterns, supporting the argument by Aksnes \emph{et al.}~\cite{Aksnes2019} that such two concepts should indeed be distinct.

Researchers are frequently evaluated purely on bibliometric indicators of impact, e.g., for tenure decisions or promotions~\cite{moher2018assessing}. Our analysis points to the fact that they should also be evaluated for their ability to innovate, as the latter might provide a very different perspective on their work. Moreover, our findings suggest that researchers in the early stages of their careers are important for disruptive innovations in science, and their role should be emphasised in scientific policy-making with regards, e.g., to the distribution of resources (e.g., grants) and to career stability.

\section{Methods}
\subsection{Data}

We collect publication and citation data for Computer Science and Physics from the AMiner citation network dataset (version 12) and the Web of Science database, respectively. The AMiner dataset contains papers from the 1960s to 2020 from DBLP, ACM, MAG, and other major sources in Computer Science~\cite{Tang:08KDD}, and it contains a total of 4,894,081 papers and 45,564,149 citations between them. The AMiner dataset has been employed in several bibliometric studies~\cite{Arif2014,zeng2019increasing,Anil2020}.

For Physics papers, we extract data from the Web of Science (WOS) database. We extract the publications of a specific selection of researchers (see below) and the citation network of their publications. Overall, we gather a total of 1,619,039 papers and 12,621,175 citations between them from 1985 to 2020. Importantly, WOS does not maintain unique author identifiers. Therefore, in order to associate an author to their publications, we apply a state-of-the-art approach proposed by Caron and van Eck to disambiguate author names~\cite{caron2014large}. Specifically, this method computes a similarity score between pairs of authors based on a series of attributes, including ORCID identifiers, names, affiliations, emails, coauthors, grant numbers, subject categories, journals, self-citations, bibliographic coupling, and co-citations. The higher the similarity score between two authors, the more likely they are to be considered the same individual. A recent study has demonstrated that this method is able to outperform other unsupervised disambiguation methods~\cite{tekles2020author}, with precision and recall scores higher than 90\%.

In our analysis, we only calculate disruption scores for papers published before 2016, thereby allowing papers in our pool to accumulate citations for at least 5 years. We set filtering criteria in line with~\cite{Li2019}, selecting only researchers with long-lived careers. The selected researchers are those who began their careers between 1980 and 2000, had at least 20 years of career, published at least 10 papers, and published at least one paper in every five years period. This filtering criteria give us a total of 27,641 and 34,527 researchers in Computer Science and Physics, respectively.

\subsection{The disruption score}

In order to address our research questions, we adopt the disruption score to quantify the innovation level of each paper in our datasets. Such a metric can effectively distinguish between innovative and developmental publications, and its robustness has been validated against data from scientific papers, patents, and software products~\cite{Wu2019,Funk2017}. The key idea of this indicator is that a highly innovative paper will eclipse attention towards preceding work in the same fields, i.e., subsequent publications will tend to cite such a paper more than the references in its bibliography.

More specifically, consider a citation network built around a focal paper, its references (previous papers) and subsequent papers. The subsequent papers can be classified into those that cite only the focal paper, those that cite both the focal paper and previous papers, and those that cite only the previous papers. Let us assume that the number of papers in each group is $n_i$, $n_j$ and $n_k$, respectively. Then the disruption score is given by 
\begin{equation}
D = \frac{n_i - n_j}{n_i + n_j + n_k}
\end{equation}
where $n_i - n_j$ measures the extent to which the focal paper has eclipsed attention towards previous papers, and $n_i + n_j + n_k$ represents the total number of subsequent papers in the network.

According to the above definition, the disruption score ranges from -1 to 1. A positive score suggests that the focal paper attracts more attention from subsequent papers than its references, which means that the focal paper is more innovative. If a focal paper is innovative enough, then its disruption $D$ should be close to 1. Similarly, a negative score means that the focal paper is likely to be a developmental piece of work. The closer the score to -1, the more developmental the paper will be. Therefore, the disruption score enables us not only to quantify each paper's innovation level but also to compare the innovation level among different papers.

\section*{Acknowledgments}
G.L. acknowledges support from an EPSRC Early Career Fellowship in Digital Economy (Grant No. EP/N006062/1). We thank Ye Sun for help with author disambiguation in WoS data.

\bibliographystyle{unsrt}  
\bibliography{templateArxiv}  

\newpage
\section*{Appendix}

\subsection*{Appendix A. The non-randomness of scientific innovation: additional details }

\begin{figure}[h!]
\centering
\includegraphics[width=0.9\textwidth]{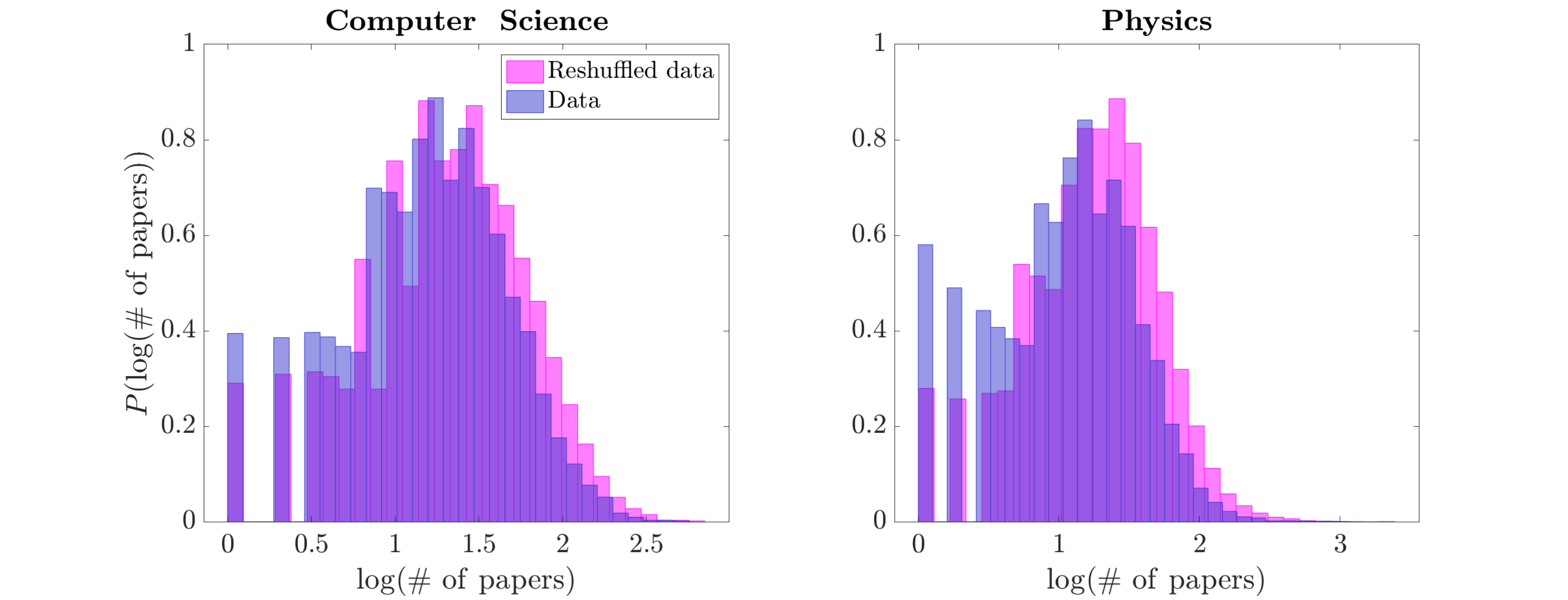}
\caption{Normalised histograms of the number of papers published by researchers before their peak year in Computer Science (left) and Physics (right) based on the original (purple) and randomised (pink) data after log transformations.}
\label{fig:papers2peak}
\end{figure}

We demonstrate the non-randomness of scientific innovation by comparing the distribution of the time taken by researchers to reach their innovation peaks based on their original and reshuffled sequences of publications. Here we further validate this result by comparing the distributions of the numbers of papers published by researchers before their peak year. The comparison of distributions is presented in Fig.~\ref{fig:papers2peak}. It can be seen that in both disciplines the distributions obtained from the original and randomized data are significantly different (in both cases $p < 0.01$, two-sided KS test).

\subsection*{Appendix B. Characterising peak year innovation: additional details}

\begin{figure}[h!]
\centering
\includegraphics[width=0.9\textwidth]{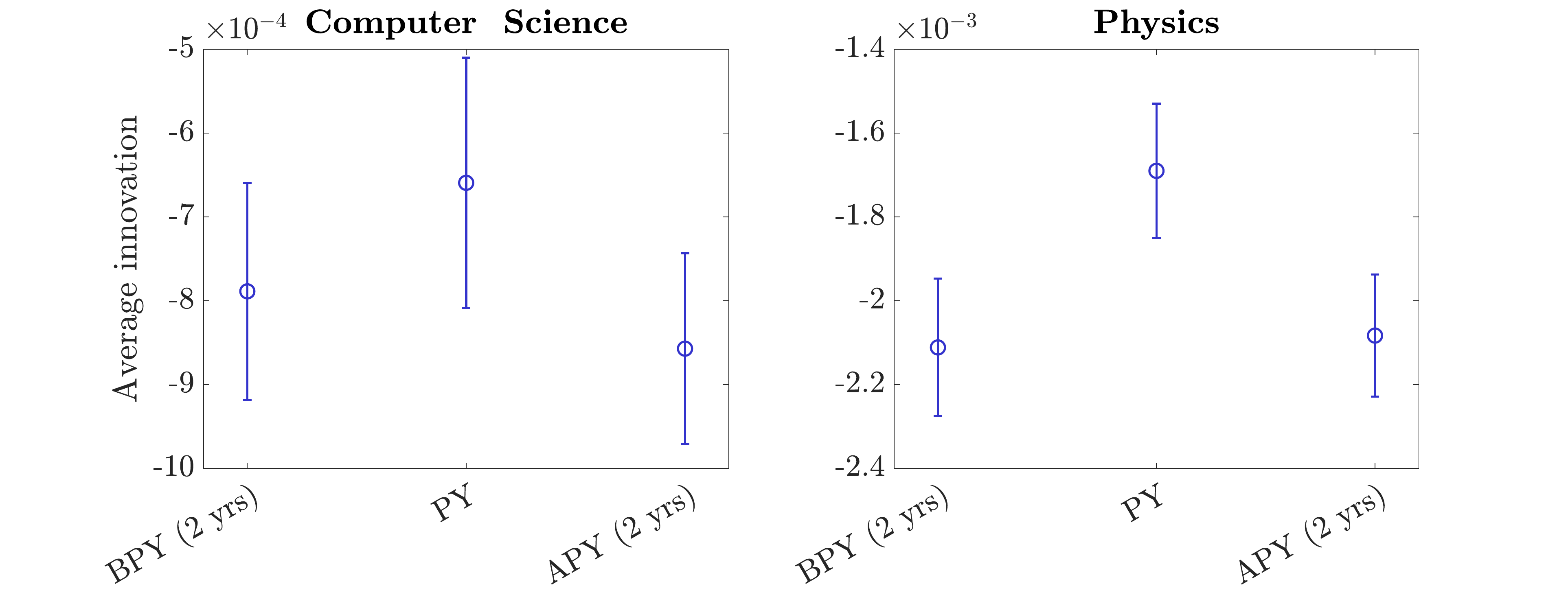}
\caption{Average innovation levels achieved by researchers with papers published 2 years before their peak year (BPY), during their peak year (excluding the paper responsible for the peak itself, PY), and 2 years after their peak year (APY). Error bars represent standard errors. In both disciplines, the average innovation level of the PY papers is higher than for papers published in other phases.}
\label{fig:innovation_levels_2yrs}
\end{figure}

In this section, we present a figure that compares innovation levels (as measured by disruption scores) of the `before peak year' (BPY), `after peak year' (APY), and the `peak year' (PY) periods. Notably, here the innovation levels of the BPY and APY phases are computed with before/after peak year periods of 2 years, while in the main text the innovation levels of BPY and APY groups are calculated with full publication histories. In this way, we mitigate potential biases due to different lengths or stages in scientific careers. For this analysis, we have in Computer Science and Physics $N_{CS} = 20,980$ and $N_{PHY} = 28,867$ for BPY, $N_{CS} = 18,640$ and $N_{PHY} = 26,543$ for PY, and $N_{CS} = 21,987$ and $N_{PHY} = 30,792$ for APY. The result is plotted in in Fig.~\ref{fig:innovation_levels_2yrs}. We find that the average disruption score is higher in the peak year than in the other two phases ($p < 0.01$ in all pairwise MWU tests and two-sided KS tests, except for $p = 0.021$ in the MWU test between PY and BPY in Computer Science), which further supports the existence of a `magical year' in scientific careers.   

\subsection*{Appendix C. Distribution plots of effort-related variables}

\begin{figure}[h!]
\centering
\includegraphics[width=1.1\textwidth]{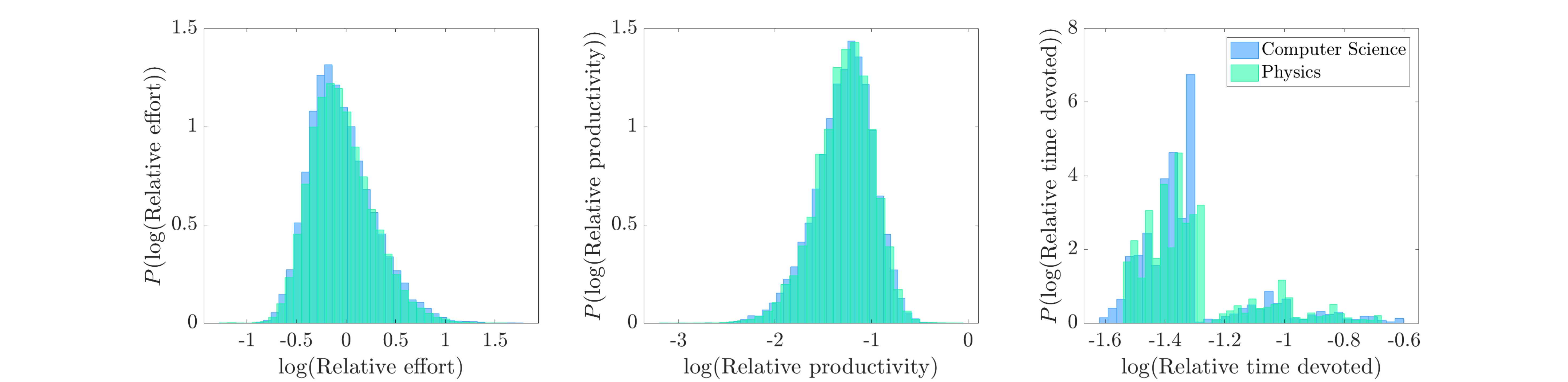}
\caption{Normalised histograms for the distributions of relative effort (left), relative productivity (center), and relative time devoted (right) associated to the peak year for researchers in our pool.}
\label{fig:distributions}
\end{figure}

To quantify the determinants of peak year innovation, we develop a series of linear models to measure the relationship between a researcher's innovation during their peak year and the relative effort, relative productivity, and relative time devotion associated to that year. Here, we show the distributions of such quantities in both Computer Science and Physics, which are plotted in Fig.~\ref{fig:distributions} as normalised histograms. As it can be seen, the distributions for the (logarithm of) relative effort and relative productivity are very similar across disciplines, whereas the distributions for relative time devoted are much more irregular and display some differences.

\subsection*{Appendix D. Additional regressions on scientific innovation and impact}

In the main text, the regression results on the innovation and impact of peak year papers are obtained from variables associated with the peak year data. As a robustness check, we build linear models on the same regressand with variables computed in both the peak year and the two years before the peak year phases. In the following, we illustrate the results of all the regression models, see Table 1-8. As can be seen in the innovation models, in both disciplines relative effort and relative time devotion contribute positively to the peak year innovation level, whereas relative productivity negatively affects the peak year innovation. In impact regressions, however, the same set of variables yield the opposite effect, which captures the difference in the mechanics of scientific innovation and impact. Similar results can be found when considering the corresponding absolute quantities. All these results support our argument about the determinants of peak year innovation.

\newpage
\begin{table}[htbp]
  \caption{Regression results for peak year innovation levels in Computer Science obtained using effort-related measures computed for the peak year.}
  \begin{adjustbox}{width=0.88\textwidth,center}
    \begin{tabular}{p{8.835em}cccccc}
    \toprule
    \toprule
    \multicolumn{1}{c}{Indep. Variables} & \multicolumn{6}{c}{Computer Science: Models for peak year innovation levels with peak year data} \\
    \midrule
    \multicolumn{1}{r}{} & Model 1 & Model 2 & Model 3 & Model 4 & Model 5 & Model 6 \\
    \midrule
    \multicolumn{1}{c}{Relative effort} & \multicolumn{1}{p{4.72em}}{0.172*** (0.007)} &       &       &       &       &  \\
    \multicolumn{1}{c}{Effort} &       & \multicolumn{1}{p{4.72em}}{0.137*** (0.007)} &       &       &       &  \\
    \multicolumn{1}{c}{Relative productivity}  &       &       & \multicolumn{1}{p{4.72em}}{-0.135*** (0.007)} &       &       &  \\
    \multicolumn{1}{c}{Productivity} &       &       &       & \multicolumn{1}{p{4.72em}}{-0.121*** (0.007)} &       &  \\
    \multicolumn{1}{c}{Relative time devoted} &       &       &       &       & \multicolumn{1}{p{4.72em}}{0.023*** (0.007)} &  \\
    \multicolumn{1}{c}{Time devoted} &       &       &       &       &       & \multicolumn{1}{p{4.72em}}{0.021*** (0.007)} \\
    \multicolumn{1}{c}{Avg. num. of coauthors}  & \multicolumn{1}{p{4.72em}}{0.108*** (0.006)} & \multicolumn{1}{p{4.72em}}{0.099*** (0.007)} & \multicolumn{1}{p{4.72em}}{0.107*** (0.007)} & \multicolumn{1}{p{4.72em}}{0.103*** (0.007)} & \multicolumn{1}{p{4.72em}}{0.117*** (0.007)} & \multicolumn{1}{p{4.72em}}{0.117*** (0.007)} \\
    \multicolumn{1}{c}{Avg. prev. innovation} & \multicolumn{1}{p{4.72em}}{0.201*** (0.006)} & \multicolumn{1}{p{4.72em}}{0.209*** (0.006)} & \multicolumn{1}{p{4.72em}}{0.205*** (0.006)} & \multicolumn{1}{p{4.72em}}{0.209*** (0.006)} & \multicolumn{1}{p{4.72em}}{0.206*** (0.007)} & \multicolumn{1}{p{4.72em}}{0.206*** (0.007)} \\
    \multicolumn{1}{c}{Peak year} & \multicolumn{1}{p{4.72em}}{-0.251*** (0.009)} & \multicolumn{1}{p{4.72em}}{-0.256*** (0.009)} & \multicolumn{1}{p{4.72em}}{-0.221*** (0.009)} & \multicolumn{1}{p{4.72em}}{-0.256*** (0.009)} & \multicolumn{1}{p{4.72em}}{-0.285*** (0.010)} & \multicolumn{1}{p{4.72em}}{-0.272*** (0.009)} \\
    \multicolumn{1}{c}{Time to peak} & \multicolumn{1}{p{4.72em}}{0.087*** (0.009)} & \multicolumn{1}{p{4.72em}}{0.067*** (0.009)} & \multicolumn{1}{p{4.72em}}{0.040*** (0.009)} & \multicolumn{1}{p{4.72em}}{0.070*** (0.009)} & \multicolumn{1}{p{4.72em}}{0.070*** (0.010)} & \multicolumn{1}{p{4.72em}}{0.057*** (0.009)} \\
    \midrule
    \multicolumn{1}{c}{$N$} & 20,980 & 20,980 & 20,980 & 20,980 & 20,980 & 20,980 \\
    \multicolumn{1}{c}{$R^2$} & 0.142 & 0.133 & 0.132 & 0.129 & 0.116 & 0.116 \\
    \bottomrule
    \bottomrule
    \end{tabular}%
  \end{adjustbox}
  \label{tab:suptable1}%
\end{table}%

\begin{table}[htbp]
  \caption{Regression results for peak year impact in Computer Science obtained using effort-related measures computed for the peak year.}
  \begin{adjustbox}{width=0.88\textwidth,center}
    \begin{tabular}{p{8.835em}cccccc}
    \toprule
    \toprule
    \multicolumn{1}{c}{Indep. Variables} & \multicolumn{6}{c}{Computer Science: Models for peak year impact with peak year data} \\
    \midrule
    \multicolumn{1}{r}{} & Model 1 & Model 2 & Model 3 & Model 4 & Model 5 & Model 6 \\
    \midrule
    \multicolumn{1}{c}{Relative effort} & \multicolumn{1}{p{4.72em}}{-0.056*** (0.007)} &       &       &       &       &  \\
    \multicolumn{1}{c}{Effort} &       & \multicolumn{1}{p{4.72em}}{-0.182*** (0.007)} &       &       &       &  \\
    \multicolumn{1}{c}{Relative productivity}  &       &       & \multicolumn{1}{p{4.72em}}{0.021*** (0.007)} &       &       &  \\
    \multicolumn{1}{c}{Productivity} &       &       &       & \multicolumn{1}{p{4.72em}}{0.370*** (0.007)} &       &  \\
    \multicolumn{1}{c}{Relative time devoted} &       &       &       &       & \multicolumn{1}{p{4.72em}}{-0.077*** (0.008)} &  \\
    \multicolumn{1}{c}{Time devoted} &       &       &       &       &       & \multicolumn{1}{p{4.72em}}{-0.056*** (0.007)} \\
    \multicolumn{1}{c}{Avg. num. of coauthors}  & \multicolumn{1}{p{4.72em}}{0.030*** (0.007)} & \multicolumn{1}{p{4.72em}}{0.051*** (0.007)} & \multicolumn{1}{p{4.72em}}{0.028*** (0.007)} & \multicolumn{1}{p{4.72em}}{0.072*** (0.007)} & \multicolumn{1}{p{4.72em}}{0.029*** (0.007)} & \multicolumn{1}{p{4.72em}}{0.028*** (0.007)} \\
    \multicolumn{1}{c}{Avg. prev. innovation} & \multicolumn{1}{p{4.72em}}{0.035*** (0.007)} & \multicolumn{1}{p{4.72em}}{0.029*** (0.007)} & \multicolumn{1}{p{4.72em}}{0.034*** (0.007)} & \multicolumn{1}{p{4.72em}}{0.023*** (0.006)} & \multicolumn{1}{p{4.72em}}{0.033*** (0.007)} & \multicolumn{1}{p{4.72em}}{0.034*** (0.007)} \\
    \multicolumn{1}{c}{Peak year} & \multicolumn{1}{p{4.72em}}{0.008\newline{}(0.010)} & \multicolumn{1}{p{4.72em}}{-0.007 (0.010)} & \multicolumn{1}{p{4.72em}}{0.008 (0.010)} & \multicolumn{1}{p{4.72em}}{-0.037*** (0.009)} & \multicolumn{1}{p{4.72em}}{0.058*** (0.011)} & \multicolumn{1}{p{4.72em}}{0.014 (0.010)} \\
    \multicolumn{1}{c}{Time to peak} & \multicolumn{1}{p{4.72em}}{0.044*** (0.010)} & \multicolumn{1}{p{4.72em}}{0.039*** (0.009)} & \multicolumn{1}{p{4.72em}}{0.056*** (0.010)} & \multicolumn{1}{p{4.72em}}{0.012 (0.009)} & \multicolumn{1}{p{4.72em}}{0.007 (0.011)} & \multicolumn{1}{p{4.72em}}{0.052*** (0.010)} \\
    \midrule
    \multicolumn{1}{c}{$N$} & 20,980 & 20,980 & 20,980 & 20,980 & 20,980 & 20,980 \\
    \multicolumn{1}{c}{$R^2$} & 0.009 & 0.038 & 0.007 & 0.134 & 0.011 & 0.009 \\
    \bottomrule
    \bottomrule
    \end{tabular}%
  \end{adjustbox}
  \label{tab:suptable2}%
\end{table}%

\begin{table}[htbp]
  \caption{Regression results for peak year innovation levels in Physics obtained using effort-related measures computed for the peak year.}
  \begin{adjustbox}{width=0.88\textwidth,center}
    \begin{tabular}{p{8.835em}cccccc}
    \toprule
    \toprule
    \multicolumn{1}{c}{Indep. Variables} & \multicolumn{6}{c}{Physics: Models for peak year innovation levels with peak year data} \\
    \midrule
    \multicolumn{1}{r}{} & Model 1 & Model 2 & Model 3 & Model 4 & Model 5 & Model 6 \\
    \midrule
    \multicolumn{1}{c}{Relative effort} & \multicolumn{1}{p{4.72em}}{0.141*** (0.006)} &       &       &       &       &  \\
    \multicolumn{1}{c}{Effort} &       & \multicolumn{1}{p{4.72em}}{0.205*** (0.006)} &       &       &       &  \\
    \multicolumn{1}{c}{Relative productivity}  &       &       & \multicolumn{1}{p{4.72em}}{-0.106*** (0.006)} &       &       &  \\
    \multicolumn{1}{c}{Productivity} &       &       &       & \multicolumn{1}{p{4.72em}}{-0.097*** (0.006)} &       &  \\
    \multicolumn{1}{c}{Relative time devoted} &       &       &       &       & \multicolumn{1}{p{4.72em}}{0.070*** (0.006)} &  \\
    \multicolumn{1}{c}{Time devoted} &       &       &       &       &       & \multicolumn{1}{p{4.72em}}{0.059*** (0.006)} \\
    \multicolumn{1}{c}{Avg. num. of coauthors}  & \multicolumn{1}{p{4.72em}}{-0.004 (0.006)} & \multicolumn{1}{p{4.72em}}{0.007 (0.006)} & \multicolumn{1}{p{4.72em}}{-0.005 (0.006)} & \multicolumn{1}{p{4.72em}}{0.005 (0.006)} & \multicolumn{1}{p{4.72em}}{0.009 (0.006)} & \multicolumn{1}{p{4.72em}}{0.008 (0.006)} \\
    \multicolumn{1}{c}{Avg. prev. innovation} & \multicolumn{1}{p{4.72em}}{0.232*** (0.006)} & \multicolumn{1}{p{4.72em}}{0.229*** (0.006)} & \multicolumn{1}{p{4.72em}}{0.233*** (0.006)} & \multicolumn{1}{p{4.72em}}{0.233*** (0.006)} & \multicolumn{1}{p{4.72em}}{0.233*** (0.006)} & \multicolumn{1}{p{4.72em}}{0.233*** (0.006)} \\
    \multicolumn{1}{c}{Peak year} & \multicolumn{1}{p{4.72em}}{-0.143*** (0.008)} & \multicolumn{1}{p{4.72em}}{-0.152*** (0.008)} & \multicolumn{1}{p{4.72em}}{-0.118*** (0.008)} & \multicolumn{1}{p{4.72em}}{-0.149*** (0.008)} & \multicolumn{1}{p{4.72em}}{-0.187*** (0.009)} & \multicolumn{1}{p{4.72em}}{-0.150*** (0.008)} \\
    \multicolumn{1}{c}{Time to peak} & \multicolumn{1}{p{4.72em}}{0.095*** (0.008)} & \multicolumn{1}{p{4.72em}}{0.098*** (0.008)} & \multicolumn{1}{p{4.72em}}{0.059*** (0.008)} & \multicolumn{1}{p{4.72em}}{0.089*** (0.008)} & \multicolumn{1}{p{4.72em}}{0.114*** (0.009)} & \multicolumn{1}{p{4.72em}}{0.080*** (0.008)} \\
    \midrule
    \multicolumn{1}{c}{$N$} & 28,867 & 28,867 & 28,867 & 28,867 & 28,867 & 28,867 \\
    \multicolumn{1}{c}{$R^2$} & 0.086 & 0.108 & 0.078 & 0.076 & 0.071 & 0.070 \\
    \bottomrule
    \bottomrule
    \end{tabular}%
  \end{adjustbox}
  \label{tab:suptable3}%
\end{table}%

\begin{table}[htbp]
  \caption{Regression results for peak year impact in Physics obtained using effort-related measures computed for the peak year.}
  \begin{adjustbox}{width=0.88\textwidth,center}
    \begin{tabular}{p{8.835em}cccccc}
    \toprule
    \toprule
    \multicolumn{1}{c}{Indep. Variables} & \multicolumn{6}{c}{Physics: Models for peak year impact with peak year data} \\
    \midrule
    \multicolumn{1}{r}{} & Model 1 & Model 2 & Model 3 & Model 4 & Model 5 & Model 6 \\
    \midrule
    \multicolumn{1}{c}{Relative effort} & \multicolumn{1}{p{4.72em}}{-0.018*** (0.006)} &       &       &       &       &  \\
    \multicolumn{1}{c}{Effort} &       & \multicolumn{1}{p{4.72em}}{-0.074*** (0.006)} &       &       &       &  \\
    \multicolumn{1}{c}{Relative productivity}  &       &       & \multicolumn{1}{p{4.72em}}{0.010 (0.006)} &       &       &  \\
    \multicolumn{1}{c}{Productivity} &       &       &       & \multicolumn{1}{p{4.72em}}{0.157*** (0.006)} &       &  \\
    \multicolumn{1}{c}{Relative time devoted} &       &       &       &       & \multicolumn{1}{p{4.72em}}{-0.038*** (0.006)} &  \\
    \multicolumn{1}{c}{Time devoted} &       &       &       &       &       & \multicolumn{1}{p{4.72em}}{-0.029*** (0.006)} \\
    \multicolumn{1}{c}{Avg. num. of coauthors}  & \multicolumn{1}{p{4.72em}}{0.006 (0.006)} & \multicolumn{1}{p{4.72em}}{0.005 (0.006)} & \multicolumn{1}{p{4.72em}}{0.006 (0.006)} & \multicolumn{1}{p{4.72em}}{0.005 (0.006)} & \multicolumn{1}{p{4.72em}}{0.003 (0.006)} & \multicolumn{1}{p{4.72em}}{0.004 (0.006)} \\
    \multicolumn{1}{c}{Avg. prev. innovation} & \multicolumn{1}{p{4.72em}}{-0.004 (0.006)} & \multicolumn{1}{p{4.72em}}{-0.003 (0.006)} & \multicolumn{1}{p{4.72em}}{-0.004 (0.006)} & \multicolumn{1}{p{4.72em}}{-0.004 (0.006)} & \multicolumn{1}{p{4.72em}}{-0.004 (0.006)} & \multicolumn{1}{p{4.72em}}{-0.004 (0.006)} \\
    \multicolumn{1}{c}{Peak year} & \multicolumn{1}{p{4.72em}}{-0.037*** (0.008)} & \multicolumn{1}{p{4.72em}}{-0.035*** (0.008)} & \multicolumn{1}{p{4.72em}}{-0.039*** (0.009)} & \multicolumn{1}{p{4.72em}}{-0.036*** (0.008)} & \multicolumn{1}{p{4.72em}}{-0.016* (0.009)} & \multicolumn{1}{p{4.72em}}{-0.036*** (0.008)} \\
    \multicolumn{1}{c}{Time to peak} & \multicolumn{1}{p{4.72em}}{0.015* (0.008)} & \multicolumn{1}{p{4.72em}}{0.010 (0.008)} & \multicolumn{1}{p{4.72em}}{0.019*** (0.008)} & \multicolumn{1}{p{4.72em}}{-0.003 (0.008)} & \multicolumn{1}{p{4.72em}}{-0.003 (0.009)} & \multicolumn{1}{p{4.72em}}{0.016* (0.008)} \\
    \midrule
    \multicolumn{1}{c}{$N$} & 28,867 & 28,867 & 28,867 & 28,867 & 28,867 & 28,867 \\
    \multicolumn{1}{c}{$R^2$} & 0.001 & 0.006 & 0.001 & 0.025 & 0.002 & 0.002 \\
    \bottomrule
    \bottomrule
    \end{tabular}%
  \end{adjustbox}
  \label{tab:suptable4}%
\end{table}%

\begin{table}[htbp]
  \caption{Regression results for peak innovation levels in Computer Science obtained using effort-related computed both for the peak year and for the two years preceding it.}
  \renewcommand{\arraystretch}{1.2} 
  \begin{adjustbox}{width=0.9\textwidth,center}
    \begin{tabular}{p{8.835em}cccccc}
    \toprule
    \toprule
    \multicolumn{1}{c}{Indep. Variables} & \multicolumn{6}{c}{Computer Science: Models on peak year innovation with all data} \\
    \midrule
    \multicolumn{1}{r}{} & Model 1 & Model 2 & Model 3 & Model 4 & Model 5 & Model 6 \\
    \midrule
    \multirow{2}[0]{*}{\textbf{Peak year variables}} &       &       &       &       &       &  \\
    \multicolumn{1}{c}{} &       &       &       &       &       &  \\
    \multirow{2}[0]{*}{Relative effort} & 0.175*** &       &       &       &       &  \\
    \multicolumn{1}{c}{} & \multicolumn{1}{c}{(0.007)} &       &       &       &       &  \\
    \multirow{2}[0]{*}{Effort} &       & 0.150*** &       &       &       &  \\
    \multicolumn{1}{c}{} &       & \multicolumn{1}{c}{(0.007)} &       &       &       &  \\
    \multirow{2}[0]{*}{Relative productivity} &       &       & -0.140*** &       &       &  \\
    \multicolumn{1}{c}{} &       &       & \multicolumn{1}{c}{(0.007)} &       &       &  \\
    \multirow{2}[0]{*}{Productivity} &       &       &       & -0.199*** &       &  \\
    \multicolumn{1}{c}{} &       &       &       & \multicolumn{1}{c}{(0.011)} &       &  \\
    \multirow{2}[0]{*}{Relative time devoted} &       &       &       &       & 0.026*** &  \\
    \multicolumn{1}{c}{} &       &       &       &       & \multicolumn{1}{c}{(0.008)} &  \\
    \multirow{2}[0]{*}{Time devoted} &       &       &       &       &       & 0.027*** \\
    \multicolumn{1}{c}{} &       &       &       &       &       & \multicolumn{1}{c}{(0.007)} \\
    \multirow{2}[0]{*}{Avg. num. of coauthors} & 0.091*** & 0.086*** & 0.086*** & 0.085*** & 0.104*** & 0.104*** \\
    \multicolumn{1}{c}{} & \multicolumn{1}{c}{(0.008)} & \multicolumn{1}{c}{(0.008)} & \multicolumn{1}{c}{(0.008)} & \multicolumn{1}{c}{(0.008)} & \multicolumn{1}{c}{(0.008)} & \multicolumn{1}{c}{(0.008)} \\
    \multirow{2}[0]{*}{\textbf{2 years before peak variables}} &       &       &       &       &       &  \\
    \multicolumn{1}{c}{} &       &       &       &       &       &  \\
    \multirow{2}[0]{*}{Relative effort} & -0.010 &       &       &       &       &  \\
    \multicolumn{1}{c}{} & \multicolumn{1}{c}{(0.007)} &       &       &       &       &  \\
    \multirow{2}[0]{*}{Effort} &       & -0.033*** &       &       &       &  \\
    \multicolumn{1}{c}{} &       & \multicolumn{1}{c}{(0.007)} &       &       &       &  \\
    \multirow{2}[0]{*}{Relative productivity } &       &       & 0.016** &       &       &  \\
    \multicolumn{1}{c}{} &       &       & \multicolumn{1}{c}{(0.007)} &       &       &  \\
    \multirow{2}[0]{*}{Productivity} &       &       &       & 0.099*** &       &  \\
    \multicolumn{1}{c}{} &       &       &       & \multicolumn{1}{c}{(0.011)} &       &  \\
    \multirow{2}[0]{*}{Relative time devoted} &       &       &       &       & 0.016** &  \\
    \multicolumn{1}{c}{} &       &       &       &       & \multicolumn{1}{c}{(0.007)} &  \\
    \multirow{2}[0]{*}{Time devoted} &       &       &       &       &       & 0.021*** \\
    \multicolumn{1}{c}{} &       &       &       &       &       & \multicolumn{1}{c}{(0.007)} \\
    \multirow{2}[0]{*}{Avg. num. of coauthors } & 0.047*** & 0.040*** & 0.055*** & 0.045*** & 0.040*** & 0.040*** \\
    \multicolumn{1}{c}{} & \multicolumn{1}{c}{(0.008)} & \multicolumn{1}{c}{(0.008)} & \multicolumn{1}{c}{(0.008)} & \multicolumn{1}{c}{(0.008)} & \multicolumn{1}{c}{(0.008)} & \multicolumn{1}{c}{(0.008)} \\
    \multirow{2}[0]{*}{\textbf{Control variables}} &       &       &       &       &       &  \\
    \multicolumn{1}{c}{} &       &       &       &       &       &  \\
    \multirow{2}[0]{*}{Avg. prev. innovation} & 0.189*** & 0.198*** & 0.192*** & 0.197*** & 0.193*** & 0.193*** \\
    \multicolumn{1}{c}{} & \multicolumn{1}{c}{(0.007)} & \multicolumn{1}{c}{(0.007)} & \multicolumn{1}{c}{(0.007)} & \multicolumn{1}{c}{(0.007)} & \multicolumn{1}{c}{(0.007)} & \multicolumn{1}{c}{(0.007)} \\
    \multirow{2}[0]{*}{Peak year} & -0.254*** & -0.261*** & -0.230*** & -0.261*** & -0.296*** & -0.273*** \\
    \multicolumn{1}{c}{} & \multicolumn{1}{c}{(0.009)} & \multicolumn{1}{c}{(0.009)} & \multicolumn{1}{c}{(0.010)} & \multicolumn{1}{c}{(0.009)} & \multicolumn{1}{c}{(0.011)} & \multicolumn{1}{c}{(0.009)} \\
    \multirow{2}[0]{*}{Time to peak} & 0.086*** & 0.065*** & 0.044*** & 0.065*** & 0.084*** & 0.062*** \\
    \multicolumn{1}{c}{} & \multicolumn{1}{c}{(0.009)} & \multicolumn{1}{c}{(0.009)} & \multicolumn{1}{c}{(0.009)} & \multicolumn{1}{c}{(0.009)} & \multicolumn{1}{c}{(0.011)} & \multicolumn{1}{c}{(0.009)} \\
    \midrule
    $N$ & 20,081 & 20,081 & 20,081 & 20,081 & 20,081 & 20,081 \\
    $R^2$ & 0.140 & 0.132 & 0.129 & 0.130 & 0.113 & 0.114 \\
    \bottomrule
    \bottomrule
    \end{tabular}%
  \end{adjustbox}
  \label{tab:suptable5}%
\end{table}%

\begin{table}[htbp]
  \caption{Regression results for peak year impact in Computer Science obtained using effort-related computed both for the peak year and for the two years preceding it.}
  \renewcommand{\arraystretch}{1.2}
  \begin{adjustbox}{width=0.9\textwidth,center}
    \begin{tabular}{p{8.835em}cccccc}
    \toprule
    \toprule
    \multicolumn{1}{c}{Indep. Variables} & \multicolumn{6}{c}{Computer Science: Models on peak year impact with all data} \\
    \midrule
    \multicolumn{1}{r}{} & Model 1 & Model 2 & Model 3 & Model 4 & Model 5 & Model 6 \\
    \midrule
    \multirow{2}[0]{*}{\textbf{Peak year variables}} &       &       &       &       &       &  \\
    \multicolumn{1}{c}{} &       &       &       &       &       &  \\
    \multirow{2}[0]{*}{Relative effort} & -0.057*** &       &       &       &       &  \\
    \multicolumn{1}{c}{} & \multicolumn{1}{c}{(0.008)} &       &       &       &       &  \\
    \multirow{2}[0]{*}{Effort} &       & -0.154*** &       &       &       &  \\
    \multicolumn{1}{c}{} &       & \multicolumn{1}{c}{(0.008)} &       &       &       &  \\
    \multirow{2}[0]{*}{Relative productivity } &       &       & 0.026*** &       &       &  \\
    \multicolumn{1}{c}{} &       &       & \multicolumn{1}{c}{(0.008)} &       &       &  \\
    \multirow{2}[0]{*}{Productivity} &       &       &       & 0.292*** &       &  \\
    \multicolumn{1}{c}{} &       &       &       & \multicolumn{1}{c}{(0.011)} &       &  \\
    \multirow{2}[0]{*}{Relative time devoted} &       &       &       &       & -0.092*** &  \\
    \multicolumn{1}{c}{} &       &       &       &       & \multicolumn{1}{c}{(0.008)} &  \\
    \multirow{2}[0]{*}{Time devoted} &       &       &       &       &       & -0.074*** \\
    \multicolumn{1}{c}{} &       &       &       &       &       & \multicolumn{1}{c}{(0.007)} \\
    \multirow{2}[0]{*}{Avg. num. of coauthors} & 0.030*** & 0.041*** & 0.030*** & 0.055*** & 0.027*** & 0.026*** \\
    \multicolumn{1}{c}{} & \multicolumn{1}{c}{(0.008)} & \multicolumn{1}{c}{(0.008)} & \multicolumn{1}{c}{(0.008)} & \multicolumn{1}{c}{(0.008)} & \multicolumn{1}{c}{(0.008)} & \multicolumn{1}{c}{(0.008)} \\
    \multirow{2}[0]{*}{\textbf{2 years before peak variables}} &       &       &       &       &       &  \\
    \multicolumn{1}{c}{} &       &       &       &       &       &  \\
    \multirow{2}[0]{*}{Relative effort} & -0.002 &       &       &       &       &  \\
    \multicolumn{1}{c}{} & \multicolumn{1}{c}{(0.008)} &       &       &       &       &  \\
    \multirow{2}[0]{*}{Effort} &       & -0.076*** &       &       &       &  \\
    \multicolumn{1}{c}{} &       & \multicolumn{1}{c}{(0.008)} &       &       &       &  \\
    \multirow{2}[0]{*}{Relative productivity } &       &       & -0.021*** &       &       &  \\
    \multicolumn{1}{c}{} &       &       & \multicolumn{1}{c}{(0.008)} &       &       &  \\
    \multirow{2}[0]{*}{Productivity} &       &       &       & 0.098*** &       &  \\
    \multicolumn{1}{c}{} &       &       &       & \multicolumn{1}{c}{(0.011)} &       &  \\
    \multirow{2}[0]{*}{Relative time devoted} &       &       &       &       & -0.073*** &  \\
    \multicolumn{1}{c}{} &       &       &       &       & \multicolumn{1}{c}{(0.008)} &  \\
    \multirow{2}[0]{*}{Time devoted} &       &       &       &       &       & -0.062*** \\
    \multicolumn{1}{c}{} &       &       &       &       &       & \multicolumn{1}{c}{(0.007)} \\
    \multirow{2}[0]{*}{Avg. num. of coauthors } & 0.0001 & 0.023*** & -0.003 & 0.031*** & 0.010 & 0.009 \\
    \multicolumn{1}{c}{} & \multicolumn{1}{c}{(0.008)} & \multicolumn{1}{c}{(0.008)} & \multicolumn{1}{c}{(0.008)} & \multicolumn{1}{c}{(0.008)} & \multicolumn{1}{c}{(0.008)} & \multicolumn{1}{c}{(0.008)} \\
    \multirow{2}[0]{*}{\textbf{Other variables}} &       &       &       &       &       &  \\
    \multicolumn{1}{c}{} &       &       &       &       &       &  \\
    \multirow{2}[0]{*}{Avg. prev. innovation} & 0.035*** & 0.027*** & 0.034*** & 0.020*** & 0.033*** & 0.034*** \\
    \multicolumn{1}{c}{} & \multicolumn{1}{c}{(0.007)} & \multicolumn{1}{c}{(0.007)} & \multicolumn{1}{c}{(0.007)} & \multicolumn{1}{c}{(0.007)} & \multicolumn{1}{c}{(0.007)} & \multicolumn{1}{c}{(0.007)} \\
    \multirow{2}[0]{*}{Peak year} & 0.005 & -0.013 & 0.011 & -0.042*** & 0.095*** & 0.008 \\
    \multicolumn{1}{c}{} & \multicolumn{1}{c}{(0.010)} & \multicolumn{1}{c}{(0.009)} & \multicolumn{1}{c}{(0.010)} & \multicolumn{1}{c}{(0.009)} & \multicolumn{1}{c}{(0.011)} & \multicolumn{1}{c}{(0.009)} \\
    \multirow{2}[0]{*}{Time to peak} & 0.039*** & 0.032*** & 0.050*** & 0.007 & -0.044*** & 0.043*** \\
    \multicolumn{1}{c}{} & \multicolumn{1}{c}{(0.010)} & \multicolumn{1}{c}{(0.009)} & \multicolumn{1}{c}{(0.009)} & \multicolumn{1}{c}{(0.009)} & \multicolumn{1}{c}{(0.012)} & \multicolumn{1}{c}{(0.009)} \\
    \midrule
    $N$ & 20,081 & 20,081 & 20,081 & 20,081 & 20,081 & 20,081 \\
    $R^2$ & 0.008 & 0.042 & 0.006 & 0.137 & 0.015 & 0.012 \\
    \bottomrule
    \bottomrule
    \end{tabular}%
  \end{adjustbox}
  \label{tab:suptable6}%
\end{table}%

\begin{table}[htbp]
  \caption{Regression results for peak year innovation level in Physics obtained using effort-related computed both for the peak year and for the two years preceding it.}
  \renewcommand{\arraystretch}{1.2}
  \begin{adjustbox}{width=0.9\textwidth,center}
    \begin{tabular}{p{8.835em}cccccc}
    \toprule
    \toprule
    \multicolumn{1}{c}{Indep. Variables} & \multicolumn{6}{c}{Physics: Models on peak year innovation with all data} \\
    \midrule
    \multicolumn{1}{r}{} & Model 1 & Model 2 & Model 3 & Model 4 & Model 5 & Model 6 \\
    \midrule
    \multirow{2}[1]{*}{\textbf{Peak year variables}} &       &       &       &       &       &  \\
    \multicolumn{1}{c}{} &       &       &       &       &       &  \\
    \multirow{2}[0]{*}{Relative effort} & 0.154*** &       &       &       &       &  \\
    \multicolumn{1}{c}{} & \multicolumn{1}{c}{(0.006)} &       &       &       &       &  \\
    \multirow{2}[0]{*}{Effort} &       & 0.203*** &       &       &       &  \\
    \multicolumn{1}{c}{} &       & \multicolumn{1}{c}{(0.006)} &       &       &       &  \\
    \multirow{2}[0]{*}{Relative productivity } &       &       & -0.105*** &       &       &  \\
    \multicolumn{1}{c}{} &       &       & \multicolumn{1}{c}{(0.006)} &       &       &  \\
    \multirow{2}[0]{*}{Productivity} &       &       &       & -0.109*** &       &  \\
    \multicolumn{1}{c}{} &       &       &       & \multicolumn{1}{c}{(0.009)} &       &  \\
    \multirow{2}[0]{*}{Relative time devoted} &       &       &       &       & 0.086*** &  \\
    \multicolumn{1}{c}{} &       &       &       &       & \multicolumn{1}{c}{(0.007)} &  \\
    \multirow{2}[0]{*}{Time devoted} &       &       &       &       &       & 0.080*** \\
    \multicolumn{1}{c}{} &       &       &       &       &       & \multicolumn{1}{c}{(0.007)} \\
    \multirow{2}[0]{*}{Avg. num. of coauthors } & -0.017** & 0.001 & -0.017** & -0.004 & 0.006 & 0.006 \\
    \multicolumn{1}{c}{} & \multicolumn{1}{c}{(0.007)} & \multicolumn{1}{c}{(0.007)} & \multicolumn{1}{c}{(0.007)} & \multicolumn{1}{c}{(0.007)} & \multicolumn{1}{c}{(0.007)} & \multicolumn{1}{c}{(0.007)} \\
    \multirow{2}[0]{*}{\textbf{2 years before peak var.}} &       &       &       &       &       &  \\
    \multicolumn{1}{c}{} &       &       &       &       &       &  \\
    \multirow{2}[0]{*}{Relative effort} & -0.027*** &       &       &       &       &  \\
    \multicolumn{1}{c}{} & \multicolumn{1}{c}{(0.007)} &       &       &       &       &  \\
    \multirow{2}[0]{*}{Effort} &       & 0.007 &       &       &       &  \\
    \multicolumn{1}{c}{} &       & \multicolumn{1}{c}{(0.006)} &       &       &       &  \\
    \multirow{2}[0]{*}{Relative productivity } &       &       & 0.004 &       &       &  \\
    \multicolumn{1}{c}{} &       &       & \multicolumn{1}{c}{(0.006)} &       &       &  \\
    \multirow{2}[0]{*}{Productivity} &       &       &       & 0.018 &       &  \\
    \multicolumn{1}{c}{} &       &       &       & \multicolumn{1}{c}{(0.010)} &       &  \\
    \multirow{2}[0]{*}{Relative time devoted} &       &       &       &       & 0.062*** &  \\
    \multicolumn{1}{c}{} &       &       &       &       & \multicolumn{1}{c}{(0.007)} &  \\
    \multirow{2}[0]{*}{Time devoted} &       &       &       &       &       & 0.056*** \\
    \multicolumn{1}{c}{} &       &       &       &       &       & \multicolumn{1}{c}{(0.007)} \\
    \multirow{2}[0]{*}{Avg. num. of coauthors } & 0.017** & 0.010 & 0.019** & 0.012 & 0.003 & 0.003 \\
    \multicolumn{1}{c}{} & \multicolumn{1}{c}{(0.007)} & \multicolumn{1}{c}{(0.007)} & \multicolumn{1}{c}{(0.007)} & \multicolumn{1}{c}{(0.007)} & \multicolumn{1}{c}{(0.007)} & \multicolumn{1}{c}{(0.007)} \\
    \multirow{2}[0]{*}{\textbf{Other variables}} &       &       &       &       &       &  \\
    \multicolumn{1}{c}{} &       &       &       &       &       &  \\
    \multirow{2}[0]{*}{Avg. prev. innovation} & 0.248*** & 0.245*** & 0.250*** & 0.250*** & 0.247*** & 0.248*** \\
    \multicolumn{1}{c}{} & \multicolumn{1}{c}{(0.006)} & \multicolumn{1}{c}{(0.006)} & \multicolumn{1}{c}{(0.006)} & \multicolumn{1}{c}{(0.006)} & \multicolumn{1}{c}{(0.006)} & \multicolumn{1}{c}{(0.006)} \\
    \multirow{2}[0]{*}{Peak year} & -0.137*** & -0.146*** & -0.114*** & -0.141*** & -0.214*** & -0.140*** \\
    \multicolumn{1}{c}{} & \multicolumn{1}{c}{(0.008)} & \multicolumn{1}{c}{(0.008)} & \multicolumn{1}{c}{(0.009)} & \multicolumn{1}{c}{(0.008)} & \multicolumn{1}{c}{(0.010)} & \multicolumn{1}{c}{(0.008)} \\
    \multirow{2}[0]{*}{Time to peak} & 0.091*** & 0.094*** & 0.062*** & 0.088*** & 0.151*** & 0.084*** \\
    \multicolumn{1}{c}{} & \multicolumn{1}{c}{(0.008)} & \multicolumn{1}{c}{(0.008)} & \multicolumn{1}{c}{(0.008)} & \multicolumn{1}{c}{(0.008)} & \multicolumn{1}{c}{(0.010)} & \multicolumn{1}{c}{(0.008)} \\
    \midrule  
    $N$ & 25,222 & 25,222 & 25,222 & 25,222 & 25,222 & 25,222 \\
    $R^2$ & 0.095 & 0.116 & 0.084 & 0.083 & 0.081 & 0.080 \\
    \bottomrule
    \bottomrule
    \end{tabular}%
  \end{adjustbox}
  \label{tab:suptable7}%
\end{table}%

\begin{table}[htbp]
  \caption{Regression results for peak year impact in Physics obtained using effort-related computed both for the peak year and for the two years preceding it.}
  \renewcommand{\arraystretch}{1.2}
  \begin{adjustbox}{width=0.9\textwidth,center}
    \begin{tabular}{p{8.835em}cccccc}
    \toprule
    \toprule
    \multicolumn{1}{c}{Indep. Variables} & \multicolumn{6}{c}{Physics: Models on peak year impact with all data} \\
    \midrule
    \multicolumn{1}{r}{} & Model 1 & Model 2 & Model 3 & Model 4 & Model 5 & Model 6 \\
    \midrule
    \multirow{2}[1]{*}{\textbf{Peak year variables}} &       &       &       &       &       &  \\
    \multicolumn{1}{c}{} &       &       &       &       &       &  \\
    \multirow{2}[0]{*}{Relative effort} & -0.034*** &       &       &       &       &  \\
    \multicolumn{1}{c}{} & \multicolumn{1}{c}{(0.007)} &       &       &       &       &  \\
    \multirow{2}[0]{*}{Effort} &       & -0.080*** &       &       &       &  \\
    \multicolumn{1}{c}{} &       & \multicolumn{1}{c}{(0.007)} &       &       &       &  \\
    \multirow{2}[0]{*}{Relative productivity} &       &       & 0.021*** &       &       &  \\
    \multicolumn{1}{c}{} &       &       & \multicolumn{1}{c}{(0.007)} &       &       &  \\
    \multirow{2}[0]{*}{Productivity} &       &       &       & 0.124*** &       &  \\
    \multicolumn{1}{c}{} &       &       &       & \multicolumn{1}{c}{(0.010)} &       &  \\
    \multirow{2}[0]{*}{Relative time devoted} &       &       &       &       & -0.057*** &  \\
    \multicolumn{1}{c}{} &       &       &       &       & \multicolumn{1}{c}{(0.007)} &  \\
    \multirow{2}[0]{*}{Time devoted} &       &       &       &       &       & -0.046*** \\
    \multicolumn{1}{c}{} &       &       &       &       &       & \multicolumn{1}{c}{(0.007)} \\
    \multirow{2}[0]{*}{Avg. num. of coauthors} & 0.014 & 0.004 & 0.014 & -0.001 & 0.005 & 0.006 \\
    \multicolumn{1}{c}{} & \multicolumn{1}{c}{(0.007)} & \multicolumn{1}{c}{(0.007)} & \multicolumn{1}{c}{(0.007)} & \multicolumn{1}{c}{(0.007)} & \multicolumn{1}{c}{(0.007)} & \multicolumn{1}{c}{(0.007)} \\
    \multirow{2}[0]{*}{\textbf{2 years before peak variables}} &       &       &       &       &       &  \\
    \multicolumn{1}{c}{} &       &       &       &       &       &  \\
    \multirow{2}[0]{*}{Relative effort} & 0.010 &       &       &       &       &  \\
    \multicolumn{1}{c}{} & \multicolumn{1}{c}{(0.007)} &       &       &       &       &  \\
    \multirow{2}[0]{*}{Effort} &       & -0.038*** &       &       &       &  \\
    \multicolumn{1}{c}{} &       & \multicolumn{1}{c}{(0.007)} &       &       &       &  \\
    \multirow{2}[0]{*}{Relative productivity } &       &       & -0.013 &       &       &  \\
    \multicolumn{1}{c}{} &       &       & \multicolumn{1}{c}{(0.007)} &       &       &  \\
    \multirow{2}[0]{*}{Productivity} &       &       &       & 0.123*** &       &  \\
    \multicolumn{1}{c}{} &       &       &       & \multicolumn{1}{c}{(0.010)} &       &  \\
    \multirow{2}[0]{*}{Relative time devoted} &       &       &       &       & -0.034*** &  \\
    \multicolumn{1}{c}{} &       &       &       &       & \multicolumn{1}{c}{(0.007)} &  \\
    \multirow{2}[0]{*}{time devoted} &       &       &       &       &       & -0.026*** \\
    \multicolumn{1}{c}{} &       &       &       &       &       & \multicolumn{1}{c}{(0.007)} \\
    \multirow{2}[0]{*}{Avg. num. of coauthors } & 0.0004 & 0.006 & -0.002 & 0.011 & 0.006 & 0.006 \\
    \multicolumn{1}{c}{} & \multicolumn{1}{c}{(0.007)} & \multicolumn{1}{c}{(0.007)} & \multicolumn{1}{c}{(0.008)} & \multicolumn{1}{c}{(0.007)} & \multicolumn{1}{c}{(0.007)} & \multicolumn{1}{c}{(0.007)} \\
    \multirow{2}[0]{*}{\textbf{Other variables}} &       &       &       &       &       &  \\
    \multicolumn{1}{c}{} &       &       &       &       &       &  \\
    \multirow{2}[0]{*}{Avg. prev. innovation} & -0.007 & -0.004 & -0.007 & -0.006 & -0.006 & -0.007 \\
    \multicolumn{1}{c}{} & \multicolumn{1}{c}{(0.006)} & \multicolumn{1}{c}{(0.006)} & \multicolumn{1}{c}{(0.006)} & \multicolumn{1}{c}{(0.006)} & \multicolumn{1}{c}{(0.006)} & \multicolumn{1}{c}{(0.006)} \\
    \multirow{2}[0]{*}{Peak year} & -0.038*** & -0.035*** & -0.039*** & -0.038*** & 0.008 & -0.037*** \\
    \multicolumn{1}{c}{} & \multicolumn{1}{c}{(0.009)} & \multicolumn{1}{c}{(0.009)} & \multicolumn{1}{c}{(0.009)} & \multicolumn{1}{c}{(0.008)} & \multicolumn{1}{c}{(0.010)} & \multicolumn{1}{c}{(0.009)} \\
    \multirow{2}[0]{*}{Time to peak} & 0.014 & 0.007 & 0.019** & -0.013 & -0.028*** & 0.014 \\
    \multicolumn{1}{c}{} & \multicolumn{1}{c}{(0.009)} & \multicolumn{1}{c}{(0.008)} & \multicolumn{1}{c}{(0.009)} & \multicolumn{1}{c}{(0.008)} & \multicolumn{1}{c}{(0.010)} & \multicolumn{1}{c}{(0.008)} \\
    \midrule
    $N$ & 25,222 & 25,222 & 25,222 & 25,222 & 25,222 & 25,222 \\
    $R^2$ & 0.002 & 0.011 & 0.001 & 0.054 & 0.004 & 0.003 \\
    \bottomrule
    \bottomrule
    \end{tabular}%
  \end{adjustbox}
  \label{tab:suptable8}%
\end{table}%

\end{document}